# Practitioners' Testimonials about Software Testing

Pradeep Waychal
*Guruji Education Foundation.*
*Chhatrapati Shahu Institute*
Thane, India
pradeep.waychal@gmail.com

Luiz Fernando Capretz
*Electrical & Computer Eng.*
*Western University*
London, Canada
lcapretz@uwo.ca

Jingdong Jia
*School of Software*
*Beihang University*
Beijing, China
jiajingdong@buaa.edu.cn

Daniel Varona    Yadira Lizama
*Cultureplex Laboratories*
*Western University*
London, Canada
{dvaronac, ylizama}@uwo.ca

*Abstract*—As software systems are becoming more pervasive, they are also becoming more susceptible to failures, resulting in potentially lethal combinations. Software testing is critical to preventing software failures but is, arguably, the least understood part of the software life cycle and the toughest to perform correctly. Adequate research has been carried out in both the process and technology dimensions of testing, but not in the human dimensions. This paper attempts to fill in the gap by exploring the human dimension, i.e., trying to understand the motivation of software professionals to take up and sustain testing careers. Towards that end, a survey was conducted in four countries - India, Canada, Cuba, and China - to try to understand how professional software testers perceive and value work-related factors that could influence their motivation to take up and sustain testing careers. With a sample of 220 software professionals, we observed that very few professionals are keen to take up testing careers. Some aspects of software testing, such as the learning opportunities, appear to be a common motivator across the four countries; whereas the treatment meted out to testers as second-class citizens and the complexity of the job appeared to be the most important de-motivators. This comparative study offers useful insights that can help global software industry leaders to come up with an action plan to put the software testing profession under a new light. That could increase the number of software engineers choosing testing careers, which would facilitate quality testing.

## I. INTRODUCTION

Many researchers have been investigating practices to improve the work performance of individuals. Several theories were developed and utilized to enlarge the body of knowledge about this theme and to contribute to the improvement of industrial practices. One of the key components which has an impact on the performance and productivity of individuals is the motivation to take up and sustain a job.

Hackman and Oldham [1] focused on workers' motivation as one of the main elements that can increase performance, and the interest and attractiveness of the job. Steers et al. [2] pointed out that theories of human motivation are generally concerned with factors or events that energize, sustain, and channel human behavior over time. Couger and Zawacki [3] pioneered the research on the motivation of software developers. They based their research on the Job Characteristics Theory and discovered that software engineers from all over the world exhibited similarities regarding their growth need strengths (GNS). More recently and specifically, researchers such as Beecham et al. [4], Sharp et al. [5], and Franca et al. [6] have emphasized the importance of understanding what motivates software engineers. Beecham et al. [4] presented a systematic literature review on motivation in software engineering, which can be relevant for managers and leaders in software engineering practice. Sharp et al. [5] proposed the MOCC (Motivators, Outcomes, Characteristics and Context) model, which integrates research work conducted in many different contexts, cultures, and software development settings. Franca el al. [6] identified visible signs in the workplace, of motivated software engineers.

Aspects of the job that motivate software engineers include problem solving, working to benefit others, and technical challenges, though, the literature on motivation in software engineering appears to present a conflicting and partial picture. Furthermore, surveys of motivation are often aimed at how software engineers feel about the organization, rather than their profession [7]. Although models of motivation in software engineering are reported, they do not account for the changing roles and environmental settings in which software engineers operate. The studies described above have concluded that there is no clear understanding of software engineers' jobs in general, what motivates them, how they are motivated, or the outcome and benefits of motivating software engineers.

Due to this lack of clarity and the criticality of human-centered activities in software development, human factors have been receiving more attention in the software engineering field [8]. In this context, several aspects of individual and team work in software development have been studied, in an attempt to understand the particularities of the human aspects of software engineering practice, such as work motivation of professionals [4] [6] [9], individual personality [10] [11], work behaviors [12], and many other factors that can directly impact a software project.

Some of these studies were consolidated in theories. For instance, there is a theory regarding the motivation and satisfaction of software engineers that was developed based on the analysis of years of published field studies and based on the specific traits of software engineering practice [6]. This theory was proposed to support academic and industrial practice on understating the motivation of software engineers working on different phases of software life cycle such as analysts, developers, testers, managers. However, when analyzing the particular features related to different roles involved in the software development process, some studies have identified observable differences among individuals playing those different roles [10] [13] [14].

Silva et al. [9] have discussed differences related to software professionals' experience with different work-related factors, such as motivation, satisfaction, and burnout, depending on the role and the tasks performed in this process. This evidence demonstrated the importance of investigations into human factors and work characteristics that occur not only in software engineering, but also, in each role and phase of the software development process.

Nevertheless, software engineering, particularly software testing, still lacks studies on motivation, especially the motivation to take up testing careers. Therefore, it is important to focus on phases of the software process, since there are considerable differences in the mindset and skills needed to perform different software tasks [13]. Regarding this need of studying each phase of the software development process, Kanij et al. [15] and Garousi et al. [16] have discussed the lack of evidence addressing human factors in software testing. These authors observed that the current research on this topic has focused mainly on the development of testing methodologies and tools and rarely on human factors affecting professional software testers. Santos et al. [7] and Deak et al. [14] discussed the difference of opinions among software testers, regarding software testing related factors that could impact their motivation.

Florea and Stray [17] analyzed 400 job advertisements for testers in 33 countries, out of which 64% asked for soft skills. Of the advertisements asking for soft skills, there are, on average, a request for five soft skills, 11 testing skills, and five technical skills. This study shows that companies want to hire testers who can communicate well and have analytical skills. They found that there is a significant increase in the need for openness and adaptability, independent-working and team-playing. Additionally, there are new categories of soft skills, such as having work ethics, customer-focus mindset, and the ability to work under pressure.

This paper attempts to understand motivators and de-motivators that influence the decisions of software professionals to take up and sustain software testing careers across four different countries, i.e. Canada, China, Cuba, and India. The research question can be framed as "How many software professionals across different geographies are keen to take up testing careers, and what are the reasons for their choices?" Towards that, we developed a cross-sectional but simple survey-based instrument, which is presented in Appendix A. In this study we investigated how software testers perceived and valued what they do and their environmental settings. This study pointed out the importance of visualizing software testing activities as a set of human-dependent tasks and emphasized the need for research that examines critically individual assessments of software testers about software testing activities. This effort can help global industry leaders to understand the impact of work-related factors on the motivation of testing professionals to take up and sustain testing careers, which can inform and support management and leadership in this context.

## II. BACKGROUND

This section presents the theoretical background that supports this study, as well as related works in a context similar to that of this research. Despite the importance of the software industry, only a handful of studies have been done on motivating software engineers to take up and sustain testing careers.

For a long time, the term *motivation* was used as a synonym for *job satisfaction* and to describe several distinct behaviors of software engineers [6]. This satisfaction/motivation disagreement among concepts represented a problem both for academic research and industrial practice, due to the need for the proper management of motivation in software companies, to achieve higher levels of productivity among professionals at work [18], and motivating software engineers continues to be a challenging task [19]. Motivational aspects have been studied in the field, including the need to identify with the task in hand, employee participation/involvement, good management, career path, sense of belonging, rewards and incentives, etc. Just like any profession in the world, software engineers also have their own de-motivators, such as the lack of feedback from supervisors, insufficient salary, lack of growth opportunities, etc. [4].

Couger and Zawacki [3], and França et al. [6] observed that the existing theories developed in various contexts and commonly discussed in the literature, might not be completely applicable to software development environments, because individuals working with software development are part of a distinctive group of workers. Therefore, "what motivates software engineers is likely to be different from what motivates the population in general". There have been early attempts to develop software-specific motivational theories, by Franca and Silva [18], and its applications to testing [19].

Considering the importance of software testing to the development of high quality and reliable software systems [20] [21] [22], and inadequate empirical evidence about the human factors affecting this activity [16], we decided to conduct a survey to try to find out *what and how work-related factors motivate or demotivate software professionals to take up and sustain testing careers?*

Towards that, we found many researchers discussing the importance of human aspects during testing [11] [23] [24] [25] [26]. Some researchers were interested not only in understanding how these professionals feel about their work, but how and why they chose this specific career [27] [28].

Rehman et al. [29] speculate that software managers cannot motivate their subordinates because they are promoted to managerial positions based on their technical expertise, not on their soft skills Thus, when they become leaders, they do not know how to cope with human behavior issues and are not trained on how to motivate their subordinates.

Weyuker et al. [27] observed that the most skilled software testers were accustomed to changing jobs in their companies and becoming programmers, analysts, or system architects, because a career in software testing was not considered advantageous enough for most of the professionals. This scenario provides a wide variety of interpretations and raises questions such as: "how demotivated does a software tester need to be to abandon their career and follow another path in the software development process?" To answer this question, it is important to consider that motivation is an antecedent of satisfaction [6], which has a strong co-relation to job burnout, one of the main factors that can lead individuals to turnover [9].

Santos et al. [7], Waychal and Capretz [28], Deak et al. [30], Deak et al. [14], and Fernández-Sanz et al. [31] have also investigated the reasons for the lack of interest of engineering graduates and professionals in testing careers. We have discussed some of their findings in the subsequent sections.

So far, previous studies related to motivation in software testing confirm that individuals working, or expecting to work, as software testers, consider a set of elements strongly related to the type of work being performed as important. However, the evidence gathered so far does not demonstrate the level of importance that software testers attribute to each factor. The present study aims to contribute to the discussion about motivation to take up and sustain testing careers, by adding new evidence to the body of knowledge of this specific topic, using an industrial survey-based approach to collect, analyze, and synthesize opinions from software professionals.

## III. METHODOLOGY

Our study analyzed the opinions of software engineering professionals about testing careers gathered by asking a sample of professionals if they would like to choose testing careers and what they felt were the PROs and CONs of testing careers. The research method is outlined in Figure 1.

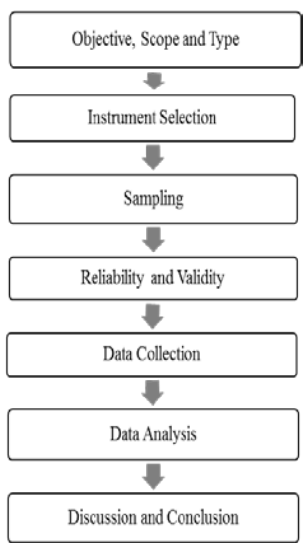

Figure 1: Research method

### A. Objective, Scope and Type

While software engineering is delivering an unprecedented performance-to-cost ratio, it is also facing tough questions about glaring failures that have caused losses of billions of dollars and human lives. Software architects, designers, and developers have been trying their best to reduce defects and require testers at various developmental stages to uncover defects before their progression to the next stages, including releasing their products to the end-users. However, very few bright individuals, across the globe, voluntarily choose testing careers, which robs the industry of good testing and of the delivery of good quality products. To change this situation, it is necessary to analyze the reasons for such apathy towards testing careers.

Our study analyzed the reasons that software engineering professionals, across four countries of not choosing testing careers. We asked 220 professionals, if they would choose testing careers, and what they saw as the PROs and CONs of these careers. The research is cross-sectional, and mixed. We did not study the event over time but at a cross-section, making the study cross-sectional. We used a qualitative method by asking for open-ended responses to the PROs and CONs of the testing career and a categorical answer about choosing a testing career, making the study a mixed one.

### B. Instrument Selection

The instrument was designed to collect responses on the motivation of software professionals to work as software testers, and to understand work-related factors in the specific context of software testing. Specifically, we asked professionals for the probability that they would choose testing careers by offering multiple choices: "Certainly Yes," "Yes," "Maybe," "No," and "Certainly Not" Since there has been limited prior research in the area, we decided to use a qualitative approach to investigate and understand phenomena within their real-life context as described by Robson [32], and asked the respondents to provide an open-ended but prioritized list of PROs and CONs, and open-ended rationale regarding their decisions on taking up testing careers.

Designing a questionnaire for a survey research requires a team of experts with research and domain expertise, which might provide both technical and practical knowledge about the topic under investigation [33]. After validation and adjustments based on the considerations received from experts, the final questionnaire was organized into three major groups of questions, written in English, and then translated into Spanish and Chinese. We strived to keep the instrument simple due to our intended use in multiple countries that have different languages and cultures. Further, when the questionnaires were translated to non-English language (i.e. Chinese and Spanish), the answers were re-translated back to English by another translator to ensure the accuracy of the translations.

Some researchers have distilled motivational factors for software tester [34] [35] [36], which could have formed the basis for our investigation. However, the three studies do not converge, indicating a possibility that some additional factors may play a role. We, therefore, designed a simple instrument without referring to the factors found in the earlier studies.

*C. Sampling*

Our sample consisted of 220 software professionals from four different countries (22 from India, 20 from Canada, 34 from China, and 144 from Cuba). While the Indian responses were sought from professionals, who were attending a testing conference, the Canadian responses were sought from alumni of a software engineering program at a university. The Chinese and Cuban responses came from professionals, who were doing part time courses at universities. We, thus, used convenience sampling both in terms of the countries as well as the professionals. Due to a varying number of respondents in the four geographies, we use percentage instead of absolute number of responses increasing the validity of comparisons.

*D. Reliability and Validity*

A reliable and valid study contains the absence of bias and higher truthfulness. The characteristics of a qualitative study are conceptualized as trustworthiness, rigor, and quality. Creswell and Miller [37] have observed that qualitative researchers employ member checking, triangulation, peer reviews, thick description, and external audits to demonstrate validity. We asked the respondents to list PROs and CONs of testing careers and the probability that they would choose a testing career, along with their rationales. We triangulated the rationale and PROs-CONs and found virtually no divergence between them. All the responses were coded by a research associate and reviewed by authors of the respective countries. Moreover, we have thickly described the responses for each category. All these measures lend the required validity to our experiment. Since the responses were not sought by supervisors of the professionals, the potential problem of evaluation apprehension was alleviated.

*E. Data Collection*

We explained the background of our study to professionals and sought their responses on their wishes to take up testing careers and on the PROs and CONs, thereof. We manually tagged all the responses and iteratively coded them until no further code changes (merging or splitting) were possible. The next subsection presents the chances of software engineering professionals taking up testing careers, and the following subsections present PROs and CONs of all the respondents. We have included the PROs and CONs that were provided by at least 5% of professionals of the respective countries. Since we excluded PROs that were chosen by less than 5% of professionals, the total of each column may not be 100%. We have also provided sample statements made by the respondents in italic.

*1) Chances of software engineering professionals taking up testing careers*

The percentage chances are depicted in Table I below.

*Canada*

Fifteen percent of professionals were certain to be testers. But close to half of the respondents were hesitant to take up a testing career.

*China*

While fifteen percent of professionals indicated their readiness to be testers, and one of them was quite certain about choosing a testing career, most them (59%), were unsure about choosing testing careers. Close to a quarter of the professionals were hesitant about taking up a testing career, and one of them was quite certain about not choosing testing as a career.

TABLE I. PERCENTAGE CHANCES OF SOFTWARE ENGINEERING PROFESSIONALS TAKING UP TESTING CAREERS

| Response | Canada | China | Cuba | India |
|---|---|---|---|---|
| Certainly Not | 15% | 3% | 17% | n.a. |
| No | 30% | 23% | 47% | n.a. |
| May be | 30% | 59% | 15% | n.a, |
| Yes | 10% | 12% | 16% | n.a. |
| Certainly Yes | 15% | 3% | 6% | n.a. |

*Cuba*

A relatively smaller number of Cuban professionals (15%) were undecided about their testing careers. Most of them (64%) did not want to work as testers, 17% of them were certain about it and 22% showed interest in testing careers, 6% with certainty.

*India*

We did not collect data from Indian professionals on the chances of taking up a testing career, as all of them were working in the testing profession.

*2) PROs of testing careers as perceived by professionals*

The analysis of responses to the PROs resulted in the following categories.

- Learning opportunities – Testers can learn about different products, technologies, techniques, and languages as well as domains such as retail, financial; and softer skills, due to more interactions with developers and customers. Testing activities provide the full background of a project's scope, modularization, and integration strategy in a short period of time and spans all project stages. Further, testing requires focusing on details and is a growing field. *This has sample statements such as, learn broad knowledge in different applications, improve your leadership skills, improve your communication and technical skills, and will have the chance to touch/experience more products and their use cases and learn about the software development life cycle.*

Important jobs – Testers are gatekeepers / in control, and are accountable and responsible for the product quality. In that sense, testing is an important part of the software life cycle. *This has sample statements such as certain amount of responsibility for the state of the system in Production or Live state and QA is very important role in software development. They focus on finding bugs which is different than developers.*

- Easy jobs – This refers to comments such as no need for originality, more mechanical, having well defined and easy processes, etc. This also helps professionals in

- balancing work and life. *This has sample statements such as clearly defined objectives and metrics and structured work schedule and easy work as compared to development.*
- Thinking, creative, challenging, and interesting job – This encompasses views about testing such as being challenging, creative, innovative, and requiring logical and analytical thinking. The testing job is also seen as lacking creativity and has come in as a CON. *This has sample statements such as you get to challenge the system, you get to think outside the box, and the challenge of insuring all aspects of a system are tested properly and completely and involves more logical thinking.*

  More jobs / Secured jobs / Stable jobs – This states that more testing jobs are available and due to the higher demands and lower supplies of testers, the jobs are secure and stable. *This has sample statements such as stability in employment – the job is always there and is always needed and lots of jobs out for software engineers in test and job security in large organizations.*

  More monetary benefits – Testing jobs come with good salary packages. *This has sample statements such as software engineers in test get paid as much as software developers and you get to work in a very lucrative and fast-paced industry.*
- Suitable for inducting freshers (new hires) – The testing activities provide smoother learning curve. *This has sample statements such as good startup for junior develops/new graduated students who do not have strong coding skills but have knowledge about programming and software engineering and the activities related to Testers are simple at first then gradually become more complex, this helps fresh new testers to have a smooth learning curve and specialized training.*
- Proximity to customers – Since the testers need to understand customer requirements, they have more chances to interact and be closer to customers. *This has sample statements such as tester activities are very client oriented and developers are far away from the business customers and testers enjoy their proximity.*
- Increases commitment to product quality - A periodical rotation of project team members to testing would increase team commitment to product quality. *This has sample statements such as testing helps you understand ill-effects of bad quality and increase commitment towards product quality and developers become more conscious of quality issues after a stint as a tester.*
- Good infrastructure - There are test engines and other automated tools that give software testers precise support. *This has sample statements such as testing infrastructure has been improving with better test engines and tools and it has become easy to carry out load and stress testing.*

The responses from each country were analyzed and presented below.

### Canada

Thirty-four percent of PROs referred to testing offering a good scope for learning in terms of domains, project architecture, etc., 16% of respondents ranked testing jobs as both important and easy, and more than 12 percent realized that testing has more jobs. The testing jobs were also seen as thinking jobs (7% PROs). Five percent PROs referred to testing having better monetary rewards.

### China

While thirty-six percent PROs referred to the fact that testing offers good learning opportunities, 32%, 14%, and 7 % of PROs referred to testing having easy, more, and important jobs, respectively. Five percent PROs referred to testing having better monetary rewards.

### Cuba

Almost half of the PRO responses (45%) referred to the learning opportunities offered by the testing field. Sixteen percent each referred to the suitability of testing profession for inducting newly hired employees (freshers) and proximity to customers that the testing offers. 13%, and 5% of PROs referred to testing assignments increasing commitment to product quality and having good infrastructure, such as test engines and automation tools.

### India

The PRO with the highest rating among Indian professionals was that it was a job that required thinking (37%). Its learning opportunities and importance fetched 28% PROs.

*3) CONs of testing careers as perceived by professionals*

The analysis of responses to the CONs resulted in the following categories:

Second-class citizen – This is a major factor and is commonly voiced by respondents. It includes testers not being involved in decision making, and being blamed for poor quality, while developers are rewarded for good quality, lack of support from management, such as unrealistic schedules and a scarcity of resources, etc. It also included the irregular work flow, lack of control over an unstable schedule, late involvement in the development cycle, and the struggle for recognition. *This has sample statements such as second-tier professional – testers are typically regarded as second-class citizens within the organization. They have almost "no-say" on the architecture and design of a system. They are always at the rear end of the development cycle, meaning challenged with very little remaining time to ship the product. The work is also tedious and repetitive. (People tease that monkeys can do this kind of work!) The pressure is high due to time constraint. It is typical to see test teams working overtime over multiple weekends prior to shipping of the system, performing regression tests over and over again, with multiple last-minute bug fixes from the development team. There will always be heated debates on whether defects are qualified or not – whether there are problems with setting up the test environment, whether there are problems with testers understanding the functionality of the system, etc. It is not surprising to arrange overnight stress test (hopefully automated, but with tester on call) to qualify the system for shipment first thing Monday morning in order to meet the deadline.*

Career development – Professionals believe that there is limited growth in the testing field. Some also believe that testers' jobs are less secure and that they are the first ones to lose their jobs in case of business downturns. *This has sample statements such as QA jobs are moving overseas and are often medium-term contract; hard to build career, often limited to QA Manager, not a lot of Director roles and regarded as "thank less"; and from recruiters/ hiring manager perspective, hard to transition in to another role and testing jobs that focus on manual testing is very career limiting and lack of growth in technical skills.*

Complexity / stressful / frustrating – This covers tester facing complex situations such as different versions of software, platform incompatibilities, defects not getting reproduced, and not being allowed sufficient time, but being held responsible for product quality. This also includes the fact that testers need to look at business and technology artifacts and understand many abstractions. The lack of clarity around requirements also adds to the difficulties. This also includes dealing with different versions and vendors of third party software, problems with testing tools, development environments or a weak infrastructure, and requiring more patience. *This has sample statements such as lots of abstraction is needed to have an adequate performance in the role of tester and Unexpected events may happen anytime rendering the performed tasks useless.*

Tedious, less creative, not challenging – This refers to the repetitive nature of testing and respondents have also used words such as monotonous and boring. *This has sample statements such as testing is repetitive work requiring loads of screen time and this is the "digital equivalent" of working as a laborer on a manufacturing assembly line; physically exhausting, mentally boring and more tedious.*

Missed development / no coding / learning opportunities – Some believe that they will be missing opportunities for professional development. Some testers do get to develop test automation systems. They may still consider that as different from actual development activity. Some also think that they will lose learning opportunities that are available to developers. *This has sample statements such as no ability to create new things, you're just checking other people's work and Creating software can be more exciting than testing software and we will be away from designing and developing system.*

Less monetary benefits – Some testers believe that testers' jobs do not have monetary benefits at par with developers. This has sample statements such as less compensation – typically, testers are paid less than architects and designers, both from a permanent and contractor perspective. For permanent positions, lower salary in turn means less benefits (such as pension, life insurance, disability allowance, etc.), which are usually a fixed percentage of the base salary.

Finding the mistakes of others – It is not easy to find out mistakes in others' work and present them. Some professionals believe this as a CON. *This includes statement such as sometimes team members hate you professionally due to found bugs and being seen as evaluating the work of peers, may to lead to workplace dissonance and lack of credibility, no matter one's competence.*

Detail oriented skills – Testers need to consider the details to find possible mistakes. This includes statement such as too many detail oriented skills are demanded from software testers and testing tasks particularities makes the software tester focus on details.

The survey respondents from each country were analyzed and described below.

*Canada*

Canadian professionals' most voted CONS were being treated as a second-class citizen (24%) and limited career development / job security (22%). Tediousness (17%), Missing development / limited learning opportunities (12%), complexity / challenging / stressful (10%) and lesser money (10%), was the next set of CONs.

*China*

The Chinese professionals' highest votes went to complexity / challenging / stressful (27%) and tedious (25%) nature of testing jobs. Limited career development / job security (15%) was the next in line. Missing development / learning opportunities and lesser financial benefits polled 9% each. The second-class citizen treatment (7%) also found a place in the list.

*Cuba*

For the Cuban professionals the most important CON about testing is finding the mistakes of others (23%). That is closely followed by complexity, challenging / stressful (20%) and the requirement of detail oriented skills (17%). The treatment of second-class citizens (15%), less monetary benefits (13%), and limited career development / learning opportunities (7%) also figured in the list.

*India*

Indian professionals' most important CONs were being treated as a second-class citizen (46%) and being complex / stressful / frustrating (40%). Tediousness (6%) of testing and missing development (6%) displayed in the list.

IV. DISCUSSIONS

In this section, we discuss the findings of our research. We first discuss insights gleaned from our studies followed by comparing our findings with that of previous studies. Then we discuss the implications of these results for research and practice.

*A. Insights from our Study*

In the four geographic regions surveyed, we found that testing was not a popular career option among software professionals. Canada has the highest percentage of professionals (25%) who wanted to take up testing careers. This is, perhaps, due to a generally higher unemployment rate and better networked readiness index, which measures drivers for ICT revolution, mapping into higher numbers of ICT jobs (Table II). In another study (to be published), we found many

Canadian students wanting to take up testing due to availability of more jobs in the area. Many Chinese professionals were ambivalent and that was perhaps due to a relatively lower unemployment rates. Most of the Cuban professionals and the Indian students [28] were very much against taking up testing careers, which may be due to better employment prospects as software developers.

TABLE II. EMPLOYMENT AND UNEMPLOYMENT IN THE ICT FIELD

|  | **Canada** | **China** | **Cuba** | **India** |
|---|---|---|---|---|
| *The Networked Readiness Index – 2016 Rank[1] | 10 | 83 | NA | 91 |
| Per Capita GDP Rank[2] | 24 | 82 | NA | 124 |
| Unemployment rate%[3] | 5.9% | 4.1% | 3.3% | 3.4% |

[1]http://online.wsj.com/public/resources/documents/GITR2016.pdf
[2]https://knoema.com/sijweyg/world-gdp-per-capita-ranking-2017-data-and-charts-forecast
[3]https://en.wikipedia.org/wiki/List_of_countries_by_unemployment_rate
*Measures the drivers of the ICT revolution

Testing offers tremendous learning opportunities as reported by professionals across the four countries; barring Indian professionals, whose most voted PRO for testing being thinking jobs, professionals from the other three countries voted that as the most common PRO. Indians professionals voted that as the second PRO. The Chinese professionals' second PRO, on the other hand, was easiness of jobs. Except Cubans, other professionals also viewed the importance of testing jobs as another PRO. The Cuban PROs were, barring learning opportunities, different and included the suitability of testing jobs for inducting freshers, proximity to customers, and an increase in commitment to software quality. We need to further investigate the reasons for such differences from the Cuban contingent.

The most common de-motivators appeared to be the second-class citizen treatment meted out to the testers and complexity resulting in stress and frustration. Barring Indian professionals, others have concerns about career development and monetary benefits in testing tracks, and barring Cuban professionals, others were concerned about tediousness and missing development aspects of testing careers. Cuban professionals had different views and pointed out difficulties in finding mistakes of others, requirement of detail oriented skills as CONs of testing career. In fact, the "finding mistakes" reason was the most voted CON by the Cuban professionals.

## V. Conclusions

We developed a survey-based instrument to understand issues related to the motivation of software professionals to take up and sustain testing careers. We collected opinions from a sample of 220 software professionals from four different countries about the characterization of the work in software testing and about the factors related to the work that influence the motivation of these professionals.

The general empirical findings on motivation to take up and continue with testing careers suggest that learning opportunities and importance of jobs appear to be common motivators across the four countries. The treatment of testing professionals as second-class citizen and complexities resulting in stressful and frustrating situations appear to be common de-motivators. Some factors were identified by many survey respondents, which increases the possibility of transferability of the factors to other situations. The findings reasonably match with earlier studies that deal with motivation in testing jobs. Since we are studying the chances of taking up testing careers, monetary rewards and jobs availability have also appeared in the list.

The overall conclusion of this study is that testing jobs remain unattractive across the four countries. Nevertheless, it is heartening to see that there are more work-related factors such as challenging job, important job, than environmental / hygiene factors such as less monetary rewards, that relate to intrinsic motivation i.e. "the inherent tendency to seek out novelty and challenges, to extend and exercise one's capacities, to explore, and to learn" [38]. Wallgren and Hanse [39], in their study of Swedish IT consultants, also found that task parameters contributed more strongly to motivation than monetary incentives or company norms. However, it is important that professionals feel valued and respected and not treated as second-class citizens, which has emerged as the strongest de-motivator.

Regarding our sample, we understand that for future generalizations we need to increase the number of participants looking for more variation of gender, experience with testing approaches (methodology, automation), the type of software and organizational contexts (product companies, service companies, testing only companies) and geographies (covering Europe and the pacific rim including Japan). Moreover, qualitative studies are difficult to replicate, but can help in understanding similar cases and situations.

However, the evidence collected and synthesized so far is a starting point for discussions regarding the motivation to take up and continue with software testing roles. Motivation is not a static construct and a longitudinal study can provide better insights into the phenomenon, which is a future direction. Despite these limitations, we expect that this study offers useful insights for industrial practice to support global managers and leaders who face problems with recruiting and motivating software testers.

Acknowledgment

We are also very grateful to all participants for dedicating their time and attention to our study.

Appendix A – Survey Questions

1. What are the three PROs (in the order of importance) for taking up testing career?
   a)
   b)
   c)

2. What are three CONs (in the order of importance) for taking up testing career?
   a)
   b)
   c)

3. What are chances of my taking up testing career?
Certainly Not   No   Maybe   Yes   Certainly Yes
Reasons: